\documentclass[aps,notitlepage,superscriptaddress]{revtex4-1}

\usepackage{amsfonts,amsmath,amssymb,diagbox,dsfont,graphicx,hyperref,lmodern,multirow,xcolor}
\usepackage[UKenglish]{babel}
\usepackage[capitalize,nameinlink]{cleveref}
\usepackage[T1]{fontenc}
\usepackage[utf8]{inputenc}
\usepackage[caption=false]{subfig}

\definecolor{darkred}{rgb}{0.5,0,0}
\definecolor{darkgreen}{rgb}{0,0.5,0}
\definecolor{darkblue}{rgb}{0,0,0.5}
\hypersetup{colorlinks,breaklinks,linkcolor=darkred,citecolor=darkgreen,urlcolor=darkblue}

\DeclareMathOperator{\Tr}{Tr}

\renewcommand{\arraystretch}{1.4}
\renewcommand{\geq}{\geqslant}
\renewcommand{\leq}{\leqslant}

\newcommand{\id}{\mathds{1}}
\newcommand{\ird}{\eta^\md}
\newcommand{\irr}{\eta^\mr}
\newcommand{\irp}{\eta^\pp}
\newcommand{\irjm}{\eta^\mjm}
\newcommand{\irg}{\eta^\mg}
\newcommand{\md}{\mathrm{d}}
\newcommand{\mr}{\mathrm{r}}
\newcommand{\pp}{\mathrm{p}}
\newcommand{\mjm}{\mathrm{jm}}
\newcommand{\mg}{\mathrm{g}}
\newcommand{\Pj}[1]{P_{j_{#1}|#1}}
\newcommand{\Qj}[1]{Q_{j_{#1}|#1}}
\newcommand{\SR}{\mathrm{SR}}
\newcommand{\vj}{{\vec{\jmath}}}

\begin{document}

\author{Sébastien Designolle}
\email{sebastien.designolle@inria.fr}
\affiliation{Zuse Institute Berlin, Takustraße 7, 14195 Berlin, Germany}
\affiliation{Inria, ENS de Lyon, UCBL, LIP, 69342, Lyon Cedex 07, France}

\title{Most incompatible measurements and sum-of-squares optimisation}
\date{17th June 2026}

\begin{abstract}
  Measurement incompatibility, or joint measurability, is a cornerstone of quantum theory and a useful resource.
  For finite-dimensional systems, quantifying this resource and establishing universal bounds valid for all measurements is a long-standing problem.
  In this work, we exhibit analytical universal parent measurements giving access to bounds that beat the state of the art.
  In particular, we can show that, for relevant robustnesses, sets of anticommuting observables give rise to the most incompatible dichotomic measurements.
  We also formalise the construction of such universal parent measurements in the framework of sum-of-squares optimisation and obtain preliminary numerical results demonstrating the power of the method by improving on our own analytical values.
  All results find direct application for demonstrating genuine high-dimensional steering, that is, certifying the dimensionality of a quantum system in a one-sided device-independent manner.
\end{abstract}

\maketitle

\section{Introduction}

A defining feature of quantum theory is the existence of phenomena with no classical counterpart.
Among these, measurement incompatibility has emerged as a central concept for understanding nonclassical behaviour~\cite{GHK+23}.
Incompatibility captures the impossibility of jointly realising certain quantum measurements, and as such it underlies many foundational quantum effects, including Bell nonlocality~\cite{BCP+14} and Einstein-Podolsky-Rosen steering~\cite{UCNG20}.
The connection to quantum steering is particularly strong: it is not merely qualitative but quantitative.
Indeed, the degree of incompatibility directly determines the robustness of steering experiments against noise~\cite{QVB14,UBGP15,CS16,XCG+22}.
This insight motivates the systematic search for maximally incompatible sets of measurements~\cite{BQG+17}.

Important progress has been made in the case of pairs of measurements~\cite{BHSS13}.
In~\cite{DFK19}, the authors characterised the most incompatible pairs of measurements in a fixed dimension.
Through the relation to steering, these results yield a one-sided device-independent dimension witness, which has been demonstrated experimentally~\cite{DSU+21,CCE+23}.
In a complementary direction, \cite{TFR+21} considered the scenario where the number of outcomes is fixed instead of the dimension, and identified the most incompatible pairs in that setting.
Together, these works provide a rather complete picture for measurement incompatibility when restricted to pairs, albeit from different but complementary perspectives.

Going beyond pairs is a natural and desirable goal.
Larger sets of incompatible measurements can reveal stronger forms of nonclassicality and lead to protocols that are more robust against noise and losses~\cite{QWZ+22,GCH+23,MRR25}.
Yet, results in this direction remain scarce.
Recent work has shown that random measurements are asymptotically maximally incompatible~\cite{BLN25}, but such constructions are not experimentally viable.
Other approaches to the finite-dimensional case rely on combining pairwise incompatibility information, which only partially captures the richer structure of larger sets~\cite{Des22}.
From a technical perspective, the main obstacle is the construction of a universal parent measurement valid for arbitrary sets, which hinges on proving semidefinite positivity~\cite{DW25}.

In this work, we address this challenge by developing a hierarchy based on sum-of-squares certificates that allows for the systematic construction of universal parent measurements.
This framework both unifies and extends the analytical methods of~\cite{DSFB19,DFK19,TFR+21}, while at the same time opening the door to detailed numerical investigations.
Among our analytical results, we prove that sets of dichotomic measurements arising from anticommuting observables are maximally incompatible with respect to many incompatibility measures where this question is well defined.
On the numerical side, our results on a restricted hierarchy serve as a proof of principle for the power of the method, and we expect further research to substantially improve on them.
Overall, our work establishes a pathway to certifying incompatibility in scenarios involving larger collections of measurements, thereby shedding new light on the structure of measurement incompatibility beyond the two-measurement regime.

The structure of this article reflects the way our hierarchy was developed.
We begin by presenting purely analytical constructions of universal parent measurements, expressed as positive polynomials in certain sums of the measurements involved.
These results already improve on the state of the art, and are even tight in the case of anticommuting observables.
The general hierarchy is introduced subsequently, when we observe that positivity can be certified through more general polynomials, which we then investigate numerically.

\section{Preliminaries}

We start by fixing some notations used in this work: $k$ will always be the number of measurements, $d$ the dimension of the underlying Hilbert space, and $n_1,\ldots,n_k$ the numbers of outcomes.
As for the indices, $x\in\{1,\ldots,k\}$ will label the measurements and $a\in\{1,\ldots,n_x\}$ the outcome, so that a set of measurements will be denoted by $\{\{A_{a|x}\}_a\}_x$, often abbreviated $\{A_{a|x}\}$.
Denoting by $\delta$ the Kronecker delta and $\vj=j_1,\ldots,j_k$, we can state the definition of compatibility, also called joint measurability.
A set $\{A_{a|x}\}$ of measurements is jointly measurable if it admits a parent measurement, that is, a measurement $\{G_\vj\}_\vj$ such that, for all $a$ and $x$,
\begin{equation}
  \sum\limits_\vj \delta_{j_x,a} G_\vj = A_{a|x}.
  \label{eqn:parent}
\end{equation}

From this dichotomic definition, one can further define incompatibility measures by quantifying the robustness of the incompatibility of $\{A_{a|x}\}$ when adding various types of noise, we refer to~\cite{DFK19} for the definitions and geometric interpretations of these measures.
There also exists a distance-based quantifier~\cite{TKKB23} for which some methods developed in this work may apply, although we do not address this formally.
The incompatibility measures we consider can be cast as semidefinite programming (SDP) instances~\cite[Appendix~E.1]{DFK19}; here we recall some of these definitions.
\begin{align}
  \label{eqn:ird}
  \text{Depolarising incompatibility robustness:}\quad
  \ird_{\{A_{a|x}\}}=&
  \left\{
    \begin{array}{cl}
      \max\limits_{\eta,\{G_\vj\}_\vj} &\eta \\
      \text{s.t.}
      &\eta \leq 1,\qquad G_\vj \geq 0,\quad \forall\vj\vphantom{\sum\limits_\vj},\\
      &\sum\limits_\vj \delta_{j_x,a} G_\vj = \eta A_{a|x}+(1-\eta)\Tr A_{a|x}\frac{\id}{d}\quad\forall a,x.\\
    \end{array}
  \right.\\
  \label{eqn:irr}
  \text{Random incompatibility robustness:}\quad
  \irr_{\{A_{a|x}\}}=&
  \left\{
    \begin{array}{cl}
      \max\limits_{\eta,\{G_\vj\}_\vj} &\eta \\
      \text{s.t.}
      &\eta \leq 1,\qquad G_\vj \geq 0,\quad \forall\vj\vphantom{\sum\limits_\vj},\\
      &\sum\limits_\vj \delta_{j_x,a} G_\vj = \eta A_{a|x}+(1-\eta)\frac{\id}{n_x}\quad\forall a,x.\\
    \end{array}
  \right.\\
  \label{eqn:irg}
  \text{Generalised incompatibility robustness:}\quad
  \irg_{\{A_{a|x}\}}=&
  \left\{
    \begin{array}{cl}
      \max\limits_{\eta,\{G_\vj\}_\vj} &\eta \\
      \text{s.t.}
      &\sum\limits_\vj G_\vj = \id,\qquad G_\vj \geq 0\quad\forall\vj,\\
      &\sum\limits_\vj \delta_{j_x,a} G_\vj \geq \eta A_{a|x}\quad\forall a,x.\\
    \end{array}
  \right.
\end{align}
We also recall some inequalities between these measures~\cite[Appendix~B]{DFK19}:
\begin{equation}\label{eqn:lowerirg}
  \ird_{\{A_{a|x}\}}+\frac{1-\ird_{\{A_{a|x}\}}}{d}\leq\irg_{\{A_{a|x}\}}
  \quad\text{and}\quad
  \irr_{\{A_{a|x}\}}+\frac{1-\irr_{\{A_{a|x}\}}}{\max_x n_x}\leq\irg_{\{A_{a|x}\}}
  .
\end{equation}
They simply follow from the computation of the objective function for $\irg$ using the optimal solution of $\ird$ or $\irr$.

With these definitions, the question of the most incompatible sets of measurements with respect to one of these measures (denoted by $\ast$ below) can be formally stated~\cite[Section~II.D]{DFK19}.
Given the dimension and numbers of outcomes, this question amounts to solving the optimisation
\begin{equation}
  \chi^\ast(d;n_1,\ldots,n_k) = \min_{\{A_{a|x}\}}\eta^\ast_{\{A_{a|x}\}}
\end{equation}
over corresponding sets $\{A_{a|x}\}$ of measurements.
We also define quantities that only fix either the dimension or the number of outcomes:
\begin{equation}
  H^\ast_k(d)=\inf_{n_1,\ldots,n_k}\chi^\ast(d;n_1,\ldots,n_k)
  \quad\text{and}\quad
  \chi^\ast(n_1,\ldots,n_k)=\inf_d \chi^\ast(d;n_1,\ldots,n_k).
\end{equation}
Note that this is a slight change of notation compared to~\cite{Des22}: the quantity $H^\mg_{k,d}$ defined there is denoted $H^\mg_k(d)$ here.
Finally, for convenience, we write $\chi^\ast_k(d;n)=\chi^\ast(d;n,\ldots,n)$ and $\chi^\ast_k(n)=\chi^\ast(n,\ldots,n)$ when all $k$ numbers of outcomes are identical.

\section{Previous works}

In this section we summarise some known results on the topic, sometimes rewriting them in our notation.

We start with~\cite{BQG+17}, the only published numerical results on the topic that we are aware of.
Therein, an extensive study is conducted, with a focus on qubit measurements.
Note that the authors were apparently not aware that different incompatibility measures could give rise to different answers to the question of the most incompatible measurements, so that the distinction between $\ird$ and $\irg$ is not very apparent in the text, although it was used in the code for performance.

In~\cite{HMZ16}, the symmetric approximate cloning machine from~\cite{KW99} is used to obtain
\begin{equation}
  H^\md_k(d) \geq \frac1k\left(1+\frac{k-1}{d+1}\right),\quad\text{thus}\quad H^\mg_k(d) \geq \frac1k\left(1+2\frac{k-1}{d+1}\right)
  \label{eqn:cloning}
\end{equation}
using \cref{eqn:lowerirg}.
The idea is to approximately clone the input state into $k$ copies which are then fed to the measurements.
By duality, the (depolarising) noise can be transferred to the measurements.

In~\cite{BN18}, the following result is derived in the framework of free spectrahedra:
\begin{equation}
  \chi^\mr_k(2)=\frac{1}{\sqrt{k}},
  \label{eqn:bn18}
\end{equation}
which is attained by projective measurements coming from $k$ anticommuting observables.
Also there, the dimension-dependent bound $\chi^\mr_k(d;2)\geq1/(2d)$ is derived, but we do not highlight it here, as it was later improved, see \cref{eqn:bn22} below.

In~\cite{DFK19}, universal parent POVMs are explicitly given to show that
\begin{equation}
  H^\mg_2(d)\geq \frac{d-2+\sqrt{d^2+4d-4}}{4(d-1)},
  \quad
  \chi^\mr(n_1,n_2)\geq\frac12\left(1+\frac{1}{\sqrt{n_1n_2}+1}\right),
  \quad\text{and}\quad
  H^\mg_2(d)=\frac12\left(1+\frac{1}{\sqrt{d}}\right).
  \label{eqn:dfk19}
\end{equation}

In~\cite[Proposition~6.7]{BN20}, the cloning bound is extended to the random incompatibility robustness:
\begin{equation}
  \chi^\mr(d;n_1,\ldots,n_k)\geq \frac1k\left(1+\frac{k-1}{d\max_xn_x+1}\right).
  \label{eqn:bn20}
\end{equation}

In~\cite[Appendix~B.4]{TFR+21}, the technique from~\cite{DFK19} for fixed dimension is extended to the setting for fixed number of outcomes, yielding
\begin{equation}
  \chi^\mg_2(n_1,n_2)\geq\frac12\left(1+\frac{\sqrt{n_1}+\sqrt{n_2}+2}{n_1+n_2+\sqrt{n_1}+\sqrt{n_2}}\right).
  \label{eqn:tfr21}
\end{equation}

In~\cite{BN22}, the following dimension-dependent bound is proven:
\begin{equation}
  \chi^\mr_k(d;2)\geq4^{-\lfloor d/2\rfloor}\binom{2\lfloor d/2\rfloor}{\lfloor d/2\rfloor}\sim\sqrt{\frac{2}{\pi d}}
  \label{eqn:bn22}
\end{equation}
asymptotically.

In~\cite{Des22}, the pairwise optimal bound from~\cite{DFK19} is used to obtain (loose) bounds for more than two measurements, namely,
\begin{equation}
  H^\mg_k(d)\geq\big(H^\mg_2(d)\big)^r\left[1-2\,\Big(1-H^\mg_2(d)\Big)\left(1-\frac{2^r}{k}\right)\right],
  \label{eqn:des22}
\end{equation}
where $r=\lfloor\log_2k\rfloor$.

In~\cite{BEK+25}, methods from non-commutative polynomial optimisation are combined with the analysis of the extreme points of some free spectrahedra to demonstrate that
\begin{equation}
  \chi^\mr(2,n)\leq\frac12\left(1+\frac{1}{\sqrt{n}+1}\right)\quad\text{and}\quad\chi^\mr_4(2;2)\leq\frac{2}{\sqrt{13}}.
  \label{eqn:bek25}
\end{equation}
The authors then conjecture both inequalities to be tight.
Note their notation for $\chi^\mr(d;n_1,\ldots,n_k)$ is $s_{\mathbb{C}}(d,k,(n_1,\ldots,n_k))$.

\section{Degree two}

In this section, we illustrate the main method used in this work: exhibiting universal parent measurements that give lower bounds on some incompatibility robustness that are valid for all sets of measurements, hence for the worst case as well.
This technique was introduced in~\cite{DFK19} and further exploited in~\cite{TFR+21,Des22}.
Note that, even though the universal parent measurements shown here are \emph{only} of degree two, the results already improve on the state of the art and are even sufficient to obtain tight bounds in some cases.

\subsection{Random incompatibility robustness}
\label{sec:irr}

Exhibiting a universal parent measurement as described above is a difficult task, and a first step is to restrict the class of measurements considered without loss of generality.
For $\irr$ (as well as $\irg$), we can exploit the \emph{pre-processing monotonicity} of the incompatibility measure to study only \emph{projective} measurements, up to a Naimark dilation that would not improve the robustness~\cite{DFK19}.
The results obtained this way are naturally agnostic of the dimension of the system.

Hence, let us consider $k$ projective measurements with $n$ outcomes, denoted $\{\Pj{1}\}_{j_1=1}^n,\ldots,\{\Pj{k}\}_{j_k=1}^n$.
We define
\begin{equation}\label{eqn:G2r}
  S_\vj=\sum_{x=1}^k\Pj{x}
  \quad\text{and}\quad
  \alpha_{k,n}=\frac{k}{n}\left(1-\frac{2(n-1)}{n-2+\sqrt{n^2+4(k-1)(n-1)}}\right)
  \quad\text{to construct}\quad
  G_\vj^{(2)}\propto \left(S_\vj-\alpha_{k,n}\id\right)^2,
\end{equation}
where the proportionality factor in the definition of $G_\vj^{(2)}$ is fixed by normalisation, i.e., $\sum_\vj G_\vj^{(2)}=\id$.
Note that the positivity of $G_\vj^{(2)}$ is trivial as the polynomial $(X-\alpha_{k,n})^2$ is a square.
Our goal is to find the noise associated with this $G_\vj^{(2)}$, that is, the value of $\eta$ such that we have a feasible point in \cref{eqn:irr}.
This requires finding the marginals of $G_\vj^{(2)}$ and its normalisation factor, a task that boils down to computing those of the successive powers of $S_\vj$ as $G_\vj^{(2)}\propto S_\vj^2-2\alpha_{k,n} S_\vj+\alpha_{k,n}^2\id$:
\begin{align}
  \label{eqn:S0}
  &\sum_\vj\id&&\!\!\!\!\!\!\!\!\!\!\!\!\!\!\!=n^k\id,
  &&
  \sum_\vj\delta_{j_x,a}\id&&\!\!\!\!\!\!\!\!\!\!\!\!\!\!\!=n^{k-1}\id,
  \\
  \label{eqn:S1}
  &\sum_\vj S_\vj&&\!\!\!\!\!\!\!\!\!\!\!\!\!\!\!=kn^{k-1}\id,
  &&
  \sum_\vj\delta_{j_x,a}S_\vj&&\!\!\!\!\!\!\!\!\!\!\!\!\!\!\!=n^{k-1}P_{a|x}+(k-1)n^{k-2}\id,
  \\
  \label{eqn:S2}
  &\sum_\vj S_\vj^2&&\!\!\!\!\!\!\!\!\!\!\!\!\!\!\!=k(n+k-1)n^{k-2}\id,
  &&
  \sum_\vj\delta_{j_x,a}S_\vj^2&&\!\!\!\!\!\!\!\!\!\!\!\!\!\!\!=\big(n+2(k-1)\big)n^{k-2}P_{a|x}+(k-1)(n+k-2)n^{k-3}\id.
\end{align}
With these marginals, we can compute the desired value of $\eta$ and conclude:
\begin{equation}\label{eqn:order2r}
  \eta=\frac{n-2+\sqrt{n^2+4(k-1)(n-1)}}{2k(n-1)}\leq\chi^\mr_k(n).
\end{equation}
Note that the value of $\alpha_{k,n}$ chosen initially is precisely picked to maximise the resulting value of $\eta$.
For $k=2$, this result surpasses the best known bound for $\irr$, namely, $\frac12[1+1/(n+1)]$, see~\cref{eqn:dfk19} (setting $n_1=n_2=n$).
One could in principle use the same technique to make this bound dependent on different numbers of outcomes.

\subsection{Application to anticommuting observables}
\label{sec:pauli}

Here we identify the most incompatible sets of dichotomic measurements with respect to $\irr$ and $\irg$ by showing that
\begin{equation}\label{eqn:chigk2}
  \chi^\mr_k(2)=\frac{1}{\sqrt{k}}\quad\text{and}\quad\chi^\mg_k(2)=\frac12\left(1+\frac{1}{\sqrt{k}}\right).
\end{equation}
These results were already obtained in~\cite{BN18}, the proof therein being much more geometric than ours given below.
They may give rise to a semi-device-independent certification of the number of dichotomic measurements as studied in~\cite{VPLU24}.

On the one hand, $\chi^\mr_k(2)\geq\frac{1}{\sqrt{k}}$ directly comes from \cref{eqn:order2r} when setting $n=2$, and this can be immediately extended to $\irg$ using \cref{eqn:lowerirg}.
On the other hand, anticommuting observables allow to reach these bounds; such observables are known to exist from dimension $2^{\lfloor k/2\rfloor}$ onwards~\cite{New32}.
This tightness is a direct application of the upper bounds from~\cite[Eq.~(E6)]{DFK19}, which read:
\begin{equation}
  \irr_{\{A_{a|x}\}}\leq\frac{\lambda-\sum_x\frac{1}{n_x}}{f-\sum_x\frac{1}{n_x}},
  \quad\text{and}\quad
  \irg_{\{A_{a|x}\}}\leq\frac{\lambda}{f},
  \quad\text{where}\quad
  \lambda=\max_\vj\left\|\sum_{x=1}^kA_{j_x|x}\right\|_\infty
  \quad\text{and}\quad f=\sum_{a,x}\frac{\Tr A_{a|x}^2}{d}.
\end{equation}
For projective measurements coming from $k$ anticommuting observables, we indeed have that $f=k$ and that the minimum polynomial of the operator whose spectrum defines $\lambda$ is, by anticommutation, $X^2-kX+\frac14k(k-1)$, so that $\lambda=\frac12(k+\sqrt{k})$, which concludes the proof.

As a side note, we observe that this proof extends to other incompatibility measures defined in~\cite{DFK19}, namely, $\irp$ and $\irjm$, showing that $\chi^\pp_k(2)=1/\sqrt{k}$ and $\chi^\mjm_k(2)=2/(\sqrt{k}+1)$.
As for $\ird$, the question of the most incompatible sets of dichotomic measurements is ill-posed since this measure is not monotonic under pre-processing.

\subsection{Depolarising incompatibility robustness}
\label{sec:ird}

For $\ird$ (as well as $\irg$), we can exploit the \emph{post-processing monotonicity} of the incompatibility measure to study only \emph{rank-one} measurements, which are extremal~\cite{DFK19}.
The results obtained this way are naturally agnostic of the number of outcomes.
This section follows closely \cref{sec:irr}, trading the number of outcomes for the dimension; this parallel will be generalised in \cref{sec:sos} below.

Hence, let us consider $k$ rank-one measurements in dimension $d$, denoted $\{A_{j_1|1}\}_{j_1},\ldots,\{A_{j_k|k}\}_{j_k}$.
We further write $A_{j_x|x}=(\Tr A_{j_x|x})\Qj{x}$ with $\Qj{x}$ rank-one projectors.
With this, we define
\begin{equation}\label{eqn:G2d}
  t_\vj=\prod_{x=1}^k\Tr A_{j_x|x}
  \quad\text{and}\quad
  T_\vj=\sum_{x=1}^k\Qj{x}
  \quad\text{to construct}\quad
  H_\vj^{(2)}\propto t_\vj\left(T_\vj-\alpha_{k,d}\id\right)^2,
\end{equation}
where $\alpha_{k,d}$ is defined in \cref{eqn:G2r} and the proportionality factor is again fixed by normalisation.
Our goal is to find the noise associated with this $H_\vj^{(2)}$, that is, the value of $\eta$ such that we have a feasible point in \cref{eqn:ird}.
This requires finding the marginals and the normalisation factor of $H_\vj^{(2)}\propto t_\vj T_\vj^2-2\alpha_{k,d}t_\vj T_\vj+\alpha_{k,d}^2t_\vj\id$:
\begin{align}
  \label{eqn:T0}
  &\sum_\vj t_\vj\id&&\!\!\!\!\!\!=d^k\id,
  &&
  \sum_\vj\delta_{j_x,a}t_\vj\id&&\!\!\!\!\!\!=d^{k-1}(\Tr A_{a|x})\id,
  \\
  \label{eqn:T1}
  &\sum_\vj t_\vj T_\vj&&\!\!\!\!\!\!=kd^{k-1}\id,
  &&
  \sum_\vj\delta_{j_x,a}t_\vj T_\vj&&\!\!\!\!\!\!=d^{k-1}A_{a|x}+(k-1)d^{k-2}(\Tr A_{a|x})\id,
  \\
  \label{eqn:T2}
  &\sum_\vj t_\vj T_\vj^2&&\!\!\!\!\!\!=k(d+k-1)d^{k-2}\id,
  &&
  \sum_\vj\delta_{j_x,a}t_\vj T_\vj^2&&\!\!\!\!\!\!=\big(d+2(k-1)\big)d^{k-2}A_{a|x}+(k-1)(d+k-2)d^{k-3}(\Tr A_{a|x})\id.
\end{align}
With these marginals, we can compute the desired value of $\eta$ and conclude:
\begin{equation}\label{eqn:order2d}
  \eta=\frac{d-2+\sqrt{d^2+4(k-1)(d-1)}}{2k(d-1)}\leq H^\md_k(d).
\end{equation}
For $k=2$, this result reproduces the value from~\cite[Eq.~(28)]{DFK19}.

Interestingly, for $k=3$ and $d=2$, projections onto the eigenbases of $X,Y,Z$ saturate this bound so that we have, using~\cref{eqn:lowerirg} to extend it to $\irg$,
\begin{equation}
  H^\md_3(2)=\frac{1}{\sqrt3}\quad\text{and}\quad H^\mg_3(2)=\frac12\left(1+\frac{1}{\sqrt3}\right).
\end{equation}
This was already proven in~\cite[Appendix~E.4]{DFK19}, albeit with a more obscure proof of positivity for the parent measurement exhibited therein.
Moreover, this parent measurement involved terms of degree three as the Jordan product was being used; this is in contrast with the degree two of our parent measurement.

\section{Degree three}

In this section, we keep the form of the universal parent measurement (as a polynomial in $S_\vj$) but climb one step in the degree ladder.
For a fixed number of outcomes, we extend \cref{sec:irr} and obtain tighter bounds on $\irg$.
For a fixed dimension, we extend \cref{sec:ird} and give consequences of the obtained bounds on $\irg$ for the experimental observation of genuine high-dimensional steering.

\subsection{Fixed number of outcomes}
\label{sec:G3gn}

Like in \cref{sec:irr}, we consider $k$ projective measurements with $n$ outcomes $\{\Pj{1}\}_{j_1=1}^n,\ldots,\{\Pj{k}\}_{j_k=1}^n$, extending the final result to all measurements with the pre-processing monotonicity of $\irg$.
With
\begin{equation}\label{eqn:G3gn}
  S_\vj=\sum_{x=1}^k\Pj{x}
  \quad\text{and}\quad
  \beta_{k,n}=\frac{k+n-2-\sqrt{(k-1)(n-1)+1}}{n},
  \quad\text{we construct}\quad
  G_\vj^{(3)}\propto S_\vj\left(S_\vj-\beta_{k,n}\id\right)^2,
\end{equation}
where the proportionality factor in the definition of $G_\vj^{(3)}$ is fixed by normalisation, i.e., $\sum_\vj G_\vj^{(3)}=\id$.
Note that $G_\vj^{(3)}$ is semidefinite positive since the polynomial $X(X-\beta_{k,n})^2$ takes nonnegative values on the interval containing the eigenvalues of $S_\vj$.
Our goal is to find the noise associated with this $G_\vj^{(3)}$, that is, the value of $\eta$ such that we have a feasible point in \cref{eqn:irg}.
This requires finding a lower bound on the marginals of $G_\vj^{(3)}$ and its normalisation factor, a task that boils down to computing those of the successive powers of $S_\vj$ as $G_\vj^{(3)}\propto S_\vj^3-2\beta_{k,n} S_\vj^2+\beta_{k,n}^2S_\vj$.

The linear and quadratic terms have already been tackled in \cref{eqn:S1,eqn:S2}.
The cubic term, however, requires more care.
On the one hand, its normalisation is trivial, though somewhat tedious:
\begin{equation}\label{eqn:norm3}
  \sum_\vj S_\vj^3=k\left(n^2+3(k-1)n+(k-1)(k-2)\right)n^{k-3}\id.
\end{equation}
On the other hand, its marginals involve pinched terms like $\Pj{1}\Pj{2}\Pj{1}$ which cannot be simply summed over $j_1$.
This is where the choice of the generalised incompatibility robustness $\irg$ becomes justified: since the definition of this measure only requires a lower bound on the marginals, we can handle such terms by using the inequality
\begin{equation}\label{eqn:pinched_n}
  \sum_{j_1}\Pj{1}\Pj{2}\Pj{1}\geq\frac1n\Pj{2},
\end{equation}
for which a proof can be found, for instance, in~\cite[Eq.~(B86)]{TFR+21}.
This allows us to derive
\begin{equation}
  \sum_\vj\delta_{j_x,a}S_\vj^3\geq\left(n^2+5(k-1)n+(k-1)(3k-5)\right)n^{k-3}P_{a|x}+(k-1)\left(n^2+3(k-2)n+(k-2)(k-3)\right)n^{k-4}\id,\label{eqn:marg3}
\end{equation}
so that we can use $\id\geq P_{a|x}$ to obtain
\begin{align}
  \sum_\vj\delta_{j_x,a}G_\vj^{(3)}&\geq\left((n+2(k-1)-n\beta_{k,n})^2+(k-1)(n-k-1)\right)n^{k-3}P_{a|x}\label{eqn:margG3}\\[-5pt]
                                   &\hphantom{\geq}+(k-1)\left((n+k-2-n\beta_{k,n})^2+(k-2)(n-1)\right)n^{k-4}\id\nonumber\\[5pt]
                                   &\geq\Big(n^3+6(k-1)n^2+(k-1)(6k-11)n+(k-1)(k-2)(k-3)\\[2pt]
                                   &\hphantom{\geq}-2\left(n^2+3(k-1)n+(k-1)(k-2)\right)n\beta_{k,n}+(n+k-1)n^2\beta_{k,n}^2\Big)n^{k-4}P_{a|x}.\nonumber
\end{align}
Crucially, the coefficient of $\id$ in \cref{eqn:margG3} is nonnegative.
With this, we can deduce the noise $\eta$ associated with $G_\vj^{(3)}$:
\begin{equation}\label{eqn:chigkn}
  \eta=\frac{k+n-2+\sqrt{(k-1)(n-1)+1}}{kn}\leq\chi^\mg_k(n).
\end{equation}
Note that the value of $\beta_{k,n}$ chosen initially is precisely picked to maximise the resulting value of $\eta$.
For $n=2$, we obtain again the value studied in \cref{sec:pauli}, which is no surprise as this value, being already optimal, could not be improved on.
For $k=2$, this result reproduces~\cref{eqn:tfr21} (setting $n_1=n_2=n$).

\subsection{Fixed dimension}
\label{sec:G3gd}

Like in \cref{sec:ird}, we consider $k$ rank-one measurements in dimension $d$, denoted $\{A_{j_1|1}\}_{j_1},\ldots,\{A_{j_k|k}\}_{j_k}$, extending the final result to all measurements by using the post-processing monotonicity of $\irg$.
We write again $A_{j_x|x}=(\Tr A_{j_x|x})\Qj{x}$ with $\Qj{x}$ rank-one projectors.
With this, we recall that
\begin{equation}\label{eqn:G3gd}
  t_\vj=\prod_{x=1}^k\Tr A_{j_x|x}
  \quad\text{and}\quad
  T_\vj=\sum_{x=1}^k\Qj{x}
  \quad\text{to construct}\quad
  H_\vj^{(3)}\propto t_\vj T_\vj\left(T_\vj-\beta_{k,d}\id\right)^2,
\end{equation}
where $\beta_{k,d}$ is defined in \cref{eqn:G3gn} and the proportionality factor is fixed by normalisation.
Note that $H_\vj^{(3)}$ is semidefinite positive since the polynomial $X(X-\beta_{k,d})^2$ takes nonnegative values on the interval $[0,k]$ containing the eigenvalues of $T_\vj$.
Like in \cref{sec:G3gn}, our goal is to find the noise associated with this $H_\vj^{(3)}$.
This requires finding a lower bound on the marginals of $H_\vj^{(3)}\propto t_\vj T_\vj^3-2\beta_{k,d}t_\vj T_\vj^2+\beta_{k,d}^2t_\vj T_\vj$.

The linear and quadratic terms were treated in \cref{eqn:T1,eqn:T2}.
\addtocounter{equation}{1} 
The normalisation of the cubic term remains straightforward
\begin{equation}
  \sum_\vj t_\vj T_\vj^3=k\left(d^2+3(k-1)d+(k-1)(k-2)\right)d^{k-3}\id,
\end{equation}
and for its marginals, we need to use, instead of \cref{eqn:pinched_n}, the inequality
\begin{equation}\label{eqn:pinched_d}
  \sum_{j_1}A_{j_1|1}^{1/2}A_{j_2|2}A_{j_1|1}^{1/2}\geq\frac1dA_{j_2|2},
\end{equation}
for which a proof can be found in~\cite[Sec.~III.E.3]{DFK19}.
With this, we can derive, similarly to \cref{sec:G3gn} but using $(\Tr A_{a|x})\id\geq A_{a|x}$ instead of $\id\geq P_{a|x}$:
\begin{align}
  \begin{aligned}
    \sum_\vj\delta_{j_x,a}H_\vj^{(3)}&\geq\Big(d^3+6(k-1)d^2+(k-1)(6k-11)d+(k-1)(k-2)(k-3)\\[-5pt]
                                     &\hphantom{\geq}-2\left(d^2+3(k-1)d+(k-1)(k-2)\right)d\beta_{k,d}+(d+k-1)d^2\beta_{k,d}^2\Big)d^{k-4}A_{a|x},
  \end{aligned}
\end{align}
from which we can deduce the noise $\eta$ associated with $H_\vj^{(3)}$:
\begin{equation}\label{eqn:Hgkd}
  \eta=\frac{k+d-2+\sqrt{(k-1)(d-1)+1}}{kd}\leq H^\mg_k(d).
\end{equation}
Note that the value of $\beta_{k,d}$ chosen initially is precisely picked to maximise the resulting value of $\eta$.
For $k=2$, this result reproduces~\cite[Eq.~(59)]{DFK19}, which is tight since it is saturated by pairs of mutually unbiased bases.

\subsection{Application to genuine high-dimensional steering}

Now that we have dimension-dependent bounds, we can derive semi-device-independent dimension witnesses~\cite{DSU+21}.
In this work we do not define genuine high-dimensional steering and refer to~\cite{Des22} for details.
For $d$-preparable assemblages, plugging \cref{eqn:Hgkd} into~\cite[Eq.~(9)]{Des22}, our work gives the following upper bound on the steering robustness:
\begin{equation}\label{eqn:SR}
  \SR_{\{\sigma_{a|x}\}}\leq\frac{(k-1)d^2+\left(k^2-\left(3+\sqrt{(k-1)(d-1)+1}\right)k+3\right)d-(k-1)(k-2)}{d^2+(k-3)d+(k-1)(k-2)},
\end{equation}
for which we give numerical values in \cref{tab:low_more}.
Note that \cref{eqn:SR} is symmetric by exchange of $k$ and $d$; this only appears to us to be a coincidence.

From \cref{eqn:SR} we can derive the desired inequality, namely, a lower bound on the dimension of the system that depends on the steering robustness.
Denoting $\SR=\SR_{\{\sigma_{a|x}\}}$ for conciseness, we have
\begin{equation}\label{eqn:witness}
  d\geq\frac{(1+\SR)\left(2k^2-5k+3+\sqrt{\big(1-3k(k-2)\big)\SR^2+2\big(k(k-2)(2k-1)+1\big)\SR+(k-1)^2}-(k-3)\SR\right)}{2(k-1-\SR)^2},
\end{equation}
generalising the witness~\cite[Eq.~(5)]{DSU+21} only valid for pairs of measurements.

\begin{table}[h]
  \centering
  \subfloat[][]{
    \begin{tabular}{|c|ccccc|}
      \hline
      \diagbox{$~k~$}{$~d~$} & 2        & 3            & 4            & 5            & 6            \\ \hline
      2                      & ~0.1716~ & 0.2679       & 0.3333       & 0.3820       & 0.4202       \\
      3                      & 0.2679   & \bf ~0.4432~ & \bf ~0.5695~ & \bf ~0.6667~ & \bf ~0.7448~ \\
      4                      & 0.3333   & \bf 0.5695   & \bf 0.7463   & \bf 0.8858   & \bf 1        \\
      5                      &          & 0.6667       & \bf 0.8858   & \bf 1.0622   & \bf 1.2087   \\
      6                      &          &              & 1            & \bf 1.2087   & \bf 1.3843   \\
      7                      &          &              &              & 1.3333       & \bf 1.5350   \\
      8                      &          &              &              &              & 1.6667       \\ \hline
    \end{tabular}
    \label{tab:low_more}
  }
  \hspace{20pt}
  \subfloat[][]{
    \begin{tabular}{|c|ccccc|}
      \hline
      \diagbox{$~k~$}{$~d~$} & 2        & 3            & 4            & 5            & 6            \\ \hline
      2                      & ~0.1716~ & 0.2679       & 0.3333       & 0.3820       & 0.4202       \\
      3                      & 0.2679   & \bf ~0.4376~ & \bf ~0.5576~ & \bf ~0.6493~ & \bf ~0.7230~ \\
      4                      & 0.3333   & \bf 0.5589   & \bf 0.7241   & \bf 0.8531   & \bf 0.9587   \\
      5                      &          & \bf 0.6521   & \bf 0.8551   & \bf 1.0165   & \bf 1.1496   \\ \hline
    \end{tabular}
    \label{tab:sos_more}
  }
  \caption{
    Upper bound on the steering robustness for $d$-preparable assemblages obtained with $k$ settings.
    (a) Analytical values obtained with \cref{eqn:SR}.
    The bold values indicate the cases for which we could strictly improve on the state of the art, the other values matching it; see~\cite[Table~I]{Des22} for comparison.
    For $k=2$, our result indeed reproduces~\cite[Eq.~(25)]{Des22}, while for $k=d+2$, it coincides with the cloning machine bound~\cite[Eq.~(11)]{Des22}.
    (b) Numerical values obtained by solving the SDP \eqref{eqn:sos_pinched} at level $t=3$.
    Bold values show cases where we outperform \cref{tab:low_more}.
  }
  \label{tab:sr}
\end{table}

\section{Higher orders: unveiling a hierarchy}

For higher powers of $S_\vj$, the normalisability is no longer guaranteed in general.
In this section, we take a detour by considering the special case of mutually unbiased bases (MUBs) before coming back to general measurements where we drop the specific form of the universal parent measurement considered, namely, as a polynomial in $S_\vj$.

We first investigate parent measurements of MUBs using the fourth power of $S_\vj$, improving on some lower bounds on $\ird$ obtained in~\cite{DSFB19}.
We then realise that the bounds obtained therein involve terms of higher degree combined in a way that positivity can still be easily proven, although not following the same procedure as above (polynomials in $S_\vj$).
This allows us to relax this assumption and give a general sum-of-squares hierarchy that encompasses the previous constructions.
Finally, we give a numerical proof of principle for this hierarchy by deriving universal lower bounds on $\irg$ which improve on the results of the previous sections, as shown in \cref{tab:sr}.

\subsection{Degree four and mutually unbiased bases}
\label{sec:mub}

We consider rank-one projective measurements onto $k$ MUBs and use the notations from \cref{sec:irr}, where $n=d$.
We refer to~\cite{DEBZ10} for the definition of MUBs, standard constructions, and applications.
Here we only need their defining property: $\Pj{x_1}\Pj{x_2}\Pj{x_1}=\frac1d\Pj{x_1}$ for $x_1\neq x_2$.
This property allows us to rewrite $\Pj{x_1}\Pj{x_2}\Pj{x_1}\Pj{x_2}=\frac1d\Pj{x_1}\Pj{x_2}$, so that $S_\vj^4$ can be normalised.\\[-10pt]

Equipped with this now valid higher power, we can consider parent measurements of the form
\begin{equation}\label{eqn:G4}
  G_\vj^{(4)}\propto\left(S_\vj-\gamma^{(1)}_{k,d}\id\right)^2\left(S_\vj-\gamma^{(2)}_{k,d}\id\right)^2
\end{equation}
to derive bounds on $\ird$ valid only for MUBs.
Up to degree two, everything remains unchanged with respect to \cref{sec:irr}.
For degree three, the normalisation is still given in \cref{eqn:norm3} but for the marginals, instead of the lower bound proposed in \cref{eqn:marg3}, the mutual unbiasedness yields
\begin{equation}
  \sum_\vj\delta_{j_x,a}S_\vj^3=\left(d^2+5(k-1)d+3(k-1)(k-2)\right)d^{k-3}P_{a|x}+(k-1)\left(d^2+(3k-5)d+(k-2)(k-3)\right)d^{k-4}\id.
\end{equation}
For degree four, we can derive the normalisation
\begin{equation}
  \sum_\vj S_\vj^4=k\left(d^3+6(k-1)d^2+(k-1)(6k-11)d+(k-1)(k-2)(k-3)\right)d^{k-4}\id
\end{equation}
and the marginals
\begin{equation}
  \begin{aligned}
    \sum_\vj\delta_{j_x,a}S_\vj^4&=\left(d^3+9(k-1)d^2+2(k-1)(7k-13)d+4(k-1)(k-2)(k-3)\right)d^{k-4}P_{a|x}\\
    &+(k-1)\left(d^3+3(2k-3)d^2+(k-2)(6k-13)d+(k-2)(k-3)(k-4)\right)d^{k-5}\id.
  \end{aligned}
\end{equation}
Note that these successive powers of $S_\vj$ immediately give valid parent measurements and hence lower bounds on $\ird$ (only valid for MUBs), reproducing, for $i\in[1,4]$, the values of $\eta^{(i)}$ given in~\cite[Appendix~D.1]{DSFB19}.

Going beyond monomials, as written in \cref{eqn:G4}, gives access to even better bounds.
Numerical values for small dimensions are given in \cref{tab:MUB}, where we highlight some small cases for which this approach outperforms~\cite{DSFB19}.
In general, however, this is not the case, and understanding why in the next subsection will unlock a more general shape for the universal parent measurements that are at the centre of our work.

\begin{table}[h]
  \centering
  \begin{tabular}{|c|ccccc|}
    \hline
    \diagbox{$~k~$}{$~d~$} & 2      & 3          & 4          & 5      & 6      \\ \hline
    2                      & 0.7071 & 0.6830     & 0.6667     & 0.6545 & 0.6449 \\
    3                      & 0.5774 & \bf 0.5483 & \bf 0.5269 & 0.5107 & 0.4980 \\
    4                      &        & \bf 0.4698 & \bf 0.4466 & 0.4291 & 0.4155 \\
    5                      &        &            & \bf 0.3933 & 0.3754 & 0.3616 \\
    6                      &        &            &            & 0.3369 & 0.3231 \\
    7                      &        &            &            &        & 0.2941 \\ \hline
  \end{tabular}
  \caption{
    Lower bound on the depolarising incompatibility robustness $\ird$ for $k$ MUBs in dimension $d$.
    Bold values indicate that we outperform~\cite{DSFB19}, see Table~IV therein for comparison.
  }
  \label{tab:MUB}
\end{table}

\subsection{Normalisable polynomials and semidefinite programming}
\label{sec:sos}

In~\cite{DSFB19}, the parent measurement giving the lower bound on $\ird$ for MUBs is written as a sum of projectors, guaranteeing its positivity.
In this section, we observe that a sum-of-squares approach is possible to obtain proofs of positivity that go beyond both the polynomials in $S_\vj$ studied above and the decomposition shown in~\cite{DSFB19} for MUBs.

Let us start by writing the first term of this decomposition~\cite[Eq.~(13)]{DSFB19} as a polynomial of the projectors $\Pj{x}$:
\begin{equation}\label{eqn:Gy}
  \mathcal{G}_\vj^{(1)}\propto X_\vj^\dagger X_\vj\quad\text{where}\quad X_\vj=\Pj{1}\prod_{x=2}^k\left(\id+\alpha_{x}\sqrt{d}\Pj{x}\right),
\end{equation}
where the product is expanded from left to right.
Interestingly, $\mathcal{G}_\vj^{(1)}$ contains monomials of degree up to $2k-1$, which soon grows past the limit of degree three that was explored so far for general measurements.
The difference lies in the form chosen: \cref{eqn:Gy} is no longer a (positive) polynomial in $S_\vj$ but simply a (positive) combination of \emph{normalisable} monomials in the projectors $\Pj{x}$.

We call a monomial in the projectors $\Pj{x}$ normalisable when the sum over $\vj$ can be reduced down to a multiple of the identity only using the projector property.
In other words, there exists an order of the summations over $j_1,\ldots,j_k$ leading to a full reduction of the monomial.
For instance, $\Pj{1}\Pj{2}\Pj{1}\Pj{2}$ mentioned above is not normalisable while $\Pj{1}\Pj{2}\Pj{3}\Pj{2}$ is, since summing over $j_1$ and $j_3$ yields $\Pj{2}^2=\Pj{2}$ which can be finally summed over $j_2$.
A normalisable polynomial only has normalisable monomials.

It is straightforward that all universal parent measurements presented so far in this work only involve normalisable polynomials.
In light of the more general proof of positivity of \cref{eqn:Gy}, we can now generalise the process of finding universal lower bounds on $\irg$ to an SOS optimisation problem~\cite{TPBA24}.
More precisely, for $t\geq1$, denoting $\mathrm{SOS}_{2t}$ the set of sum-of-squares polynomials of degree up to $2t$ in the projectors $\Pj{x}$, we can write

\begin{equation}\label{eqn:sos}
  \chi^\mg_k(n)\geq
  \left\{
    \begin{array}{cl}
      \max\limits_{\eta,G} &\eta \\
      \text{s.t.}
      &\sum\limits_\vj G(\Pj{1},\ldots,\Pj{k}) = \id,\quad G(\Pj{1},\ldots,\Pj{k}) \in\mathrm{SOS}_{2t}\quad\forall\vj,\\
      &\sum\limits_\vj\delta_{j_x,a}G(\Pj{1},\ldots,\Pj{k})-\eta P_{a|x}\in\mathrm{SOS}_{2t}\quad\forall a,x.\\
    \end{array}
  \right.
\end{equation}
\cref{eqn:sos} is simply obtained by rewriting \cref{eqn:irg} when fixing the form of the parent measurement as a normalisable polynomial in the projectors $\Pj{x}$.
Importantly, the polynomial $G$ has to be normalisable, which can be enforced by choosing a basis of normalisable monomials.
In the last constraint, without loss of generality, we can set $a=1$.

Note that the similarity between \cref{sec:irr,sec:ird} (also \cref{sec:G3gn,sec:G3gd}) generalises, so that bounds on $\chi^\mg_k(n)$ (for $n$-outcome measurements) can be transposed to $H^\mg_k(n)$ (for $n$-dimensional measurements) by considering rank-one measurements instead of projective ones and adding a factor $t_\vj$ in the parent measurement.

A crucial open question that numerical investigations should clarify is whether the hierarchy converges to the actual value of $\chi^\mg_k(n)$ or $H^\mg_k(d)$, that is, whether \cref{eqn:sos} becomes tight for some $t$ (or even asymptotically).
As shown in the literature and in this work, the answer is already known to be positive in some simple cases:
\begin{itemize}
\item for $k=2$ (see~\cite[Section~III.G]{DFK19} or our remark at the end of \cref{sec:G3gd}),
\item for $d=2$ and $k=3$ (see~\cite[Appendix~E.4]{DFK19} or our simplified proof in \cref{sec:ird}),
\item for $n=2$ (see~\cref{sec:irr}).
\end{itemize}
Showing such a tightness in more general cases would demonstrate the power of this hierarchy and potentially give access to a proof that projective measurements onto complete sets of MUBs in dimension $d$ are among the most incompatible sets of $d+1$ measurements in dimension $d$.
Moreover, this could provide an interesting duality between the incompatibility properties of $d$-dimensional and $d$-outcome measurements.

Computationally, however, the hierarchy quickly becomes challenging, and we do not explore it extensively in this work, only giving some preliminary results in the next subsection.
The symmetrisation technique developed in~\cite{GP24} could prove useful to investigate this question in the smallest open case, namely, $k=3$ and $n=3$.

\subsection{Pinched-marginalisable polynomials and preliminary numerical results}
\label{sec:sos_pinched}

As mentioned above, solving \cref{eqn:sos} is computationally demanding.
Here we add some restrictions on the form of $G$ to define a less general hierarchy for which we can derive numerical results, hence demonstrating the potential of \cref{eqn:sos}.
The main idea is to select only certain types of monomials for which we can easily deal with the marginals.

Consider a normalisable polynomial in the projectors $\Pj{x}$ and denote by $\Sigma$ the operation that formally transforms this polynomial into its normalisation.
For example, $\Sigma(\Pj{1})=n^{k-1}$ or $\Sigma(\Pj{1}\Pj{2}\Pj{1})=n^{k-2}$.
This operation amounts to rewriting rules, erasing one by one the factors as we sum over all indices $j_1,\ldots,j_k$ (not necessarily in this order), applying the projector identity along the way.

The restriction of this section aims at using such rewriting rules for the marginals as well, making the last constraint from \cref{eqn:sos} much simpler as it will depend only on $k$ variables, and not all $n^k$ of them.
More precisely, we define \emph{pinched-marginalisable} monomials to be such that, for all $x$, their marginals along $x$ (when summing over all indices except for $j_x$) can be rewritten into a multiple of $\Pj{x}$ using the projector rules together with \cref{eqn:pinched_n}, which we recall for convenience: $\sum_{j_1}\Pj{1}\Pj{2}\Pj{1}\geq\frac1n\Pj{2}$ (note that using \cref{eqn:pinched_n} repeatedly is permitted since $B\geq0$ implies $ABA\geq0$ for $A$ Hermitian).
We denote $\Sigma_x$ the associated transformation.
A pinched-marginalisable polynomial only has pinched-marginalisable monomials and $\Sigma_x$ extends to polynomials by linearity.
For instance, $\Sigma_1(\Pj{1}\Pj{2}\Pj{1})=n^{k-2}$ and $\Sigma_2(\Pj{1}\Pj{2}\Pj{1})=n^{k-3}$.
Here, $\Sigma_1$ does not use \cref{eqn:pinched_n} whereas $\Sigma_2$ does; since \cref{eqn:pinched_n} is an inequality, special care has to be taken to ensure the positivity of the coefficient in the polynomial.
We do not explicitly state these conditions in the following, but they should not be dropped in practice.

Now that there are only $k$ variables involved, we can state, for $t\geq1$, the restricted hierarchy:
\begin{equation}\label{eqn:sos_pinched}
  \chi^\mg_k(n)\geq
  \left\{
    \begin{array}{cl}
      \max\limits_{\eta,G} &\eta \\
      \text{s.t.}
      &\Sigma\big(G(\Pj{1},\ldots,\Pj{k})\big)=1,\quad G(\Pj{1},\ldots,\Pj{k}) \in\mathrm{SOS}_{2t},\\
      &\Sigma_x\big(G(\Pj{1},\ldots,\Pj{k})\big)\geq\eta\quad\forall x.\\
    \end{array}
  \right.
\end{equation}
For $d$-dimensional measurements, the procedure mentioned above works as well, the only difference being that \cref{eqn:pinched_d} is now used instead of \cref{eqn:pinched_n}.
Numerical results are given in \cref{tab:sos_more}.

Note that the inherent locality of our constraints, where marginals depend only on specific measurement settings $x$, suggests a natural connection to correlative sparsity~\cite{WKKM06,WML21}, a structure that could be systematically exploited to decompose the global SDP into smaller, more tractable blocks.
Moreover, the notion of normalisability identifies a subset of relevant monomials, and further exploring term-sparsity algorithms~\cite{WML21} could provide a more systematic way to automate this selection beyond the pinched constructions introduced here.

\section{Conclusion}

\begin{table}
  \centering
  \renewcommand{\arraystretch}{1.7}
  \begin{tabular}{|c|c|c|cc|c|}
    \hline
    & Parent measurement                                                      & ~Degree~ & \multicolumn{2}{c|}{~~Lower bound on $\irg$~~}                           & Reference                 \\ \hline
    & Cloning machine                                                         & $\times$ & $\frac{13}{25}$                                        & $=0.5200$       & \cref{eqn:cloning}        \\ \cline{2-6}
    & $t_\vj\big(T_\vj-\frac12\id\big)^2$                                     & 2        & $\frac12$                                              & $=0.5000$       & \cref{eqn:G2r}            \\ \cline{2-6}
    & $t_\vj T_\vj\big(T_\vj-\frac{7-\sqrt{13}}{4}\id\big)^2$                 & 3        & ~$\frac{7+\sqrt{13}}{20}$~                             & $\approx0.5303$ & \cref{eqn:G3gn}           \\ \cline{2-6}
    &                                                                         & 2        &                                                        & $\approx0.5000$ &                           \\ \cline{3-5}
    & Pinched-marginalisable                                                  & 4        &                                                        & $\approx0.5367$ & \cref{sec:sos_pinched}    \\ \cline{3-5}
    &                                                                         & 6        &                                                        & $\approx0.5391$ &                           \\ \cline{2-6}
    \parbox[t]{5mm}{\multirow{-7}{*}{\rotatebox[origin=c]{90}{Universal}}}
    & Normalisable                                                            & ?        & \multicolumn{2}{c|}{?}                                                   & \cref{sec:sos}            \\ \hline
    & ~$\big(S_\vj-\frac{3-\sqrt6}{2}\id\big)^2\big(S_\vj-\frac54\id\big)^2$~ & 4        & $\frac{3+\sqrt6}{10}$                                  & $\approx0.5449$ & \cref{eqn:G4}             \\ \cline{2-6}
    & $X_\vj^\dagger X_\vj+\text{perm.}$                                      & 9        &                                                        & $\approx0.5439$ & \cref{eqn:Gy}             \\ \cline{2-6}
    \parbox[t]{5mm}{\multirow{-3}{*}{\rotatebox[origin=c]{90}{MUBs}}}
    & $\Pi_\vj\text{ iff }\|S_\vj\|_\infty=\frac{4+\sqrt3}{2}$                & $\infty$ & $\frac{4+\sqrt3}{10}$                                  & $\approx0.5732$ & ~\cite[Eq.~(11)]{DSFB19}~ \\ \hline
  \end{tabular}
  \caption{
    Summary of different parent measurements for the case $d=4$ and $k=5$, both universal (for all sets of five four-dimensional measurements) and only for MUBs.
    When these parent measurements are polynomials in the projectors of the rank-one measurements considered, we indicate the corresponding degree.
    All values are analytical, except for the pinched-marginalisable ones, which we obtain by solving the corresponding SDP~\eqref{eqn:sos_pinched} numerically.
    Higher values could be reached (say, with degree 8) but this was out of reach with our code; also, we did not implement the generic hierarchy for normalisable parent measurements.
    Apart from the cloning machine, all universal bounds have their equivalent for four-outcome measurements, dropping the factor $t_\vj$ and replacing $T_\vj$ by $S_\vj$.
    For MUBs, the value with degree $2k-1=9$ is obtained by sums of permutations of \cref{eqn:Gy} as described in~\cite[Eq.~(13)]{DSFB19}; the fact that it is lower than the value reached above with degree four is not true in general, see \cref{tab:MUB}.
    In the last line, we give the exact value for MUBs; $\Pi_\vj$ is the eigenprojector of the maximal eigenvalue of $S_\vj$, and it is shown in~\cite[Appendix~D.1]{DSFB19} that this parent measurement is optimal and asymptotically reached by powers of $S_\vj$.
  }
  \label{tab:summary}
\end{table}

Finding the most incompatible sets of measurements can be addressed by exhibiting universal parent measurements, which we summarise in \cref{tab:summary} (together with some constructions from the literature).
On the one hand, this work has exhausted the direction where the parent measurement comes from a positive polynomial in certain sums of the measurements.
The bounds obtained this way reproduce known results for pairs of measurements and extend them nontrivially for larger sets.
Importantly, we have shown that anticommuting observables are maximally incompatible with respect to measures where this question is relevant.
On the other hand, we have outperformed this approach through an SDP hierarchy that we defined, where the positivity of the universal parent measurements and the marginal constraint are proven via sum-of-squares certificates.
This gives more freedom in the choice of the universal parent measurement and hence unlocks tighter bounds, which we certify numerically using a restricted hierarchy that involves smaller SDP instances.

The central question opened by this work is simple: can the general hierarchy be tight?
It is already true in many edge cases identified by previous works and this one, namely, when $k=2$ (pairs), $n=2$ (dichotomic), or $d=2$ (qubit).
The most natural candidates for testing this in general would be complete sets of MUBs, starting in dimension $d=3$, which could be numerically tractable.
Whether the answer is positive or negative, improving on the bounds that we set in this work will certainly contribute to making the contrast between systems of increasing dimensions more easily observable in applications where incompatibility plays a role.

\acknowledgments
We thank Armin Tavakoli for suggesting to study Pauli strings, which was the starting point of this work, Gabriel Cobucci for sharing related notes, Mateus Araújo for helping out using Moment at an early stage of the numerical experiments, Andreas Bluhm for pointing out relevant references, and Nicola d'Alessandro and Máté Farkas for useful discussions.
Research reported in this paper was supported through the Research Campus Modal funded by the German Federal Ministry of Education and Research (fund numbers 05M14ZAM, 05M20ZBM), and partly funded within the QuantERA II Programme that has received funding from the European Union's Horizon 2020 research and innovation programme under Grant Agreement No 101017733 (VERIqTAS).

\bibliography{Des26}

\end{document}